\documentclass[10pt,twocolumn,conference]{IEEEtran}

\usepackage{graphicx}
\usepackage{amsmath}
\usepackage{amssymb}
\usepackage[caption=false]{subfig}
\usepackage[noadjust]{cite}
\usepackage{float}
\usepackage{algorithm}
\usepackage{bibentry}
\usepackage{balance}
\usepackage{algorithm}
\usepackage{algorithmicx}
\usepackage{algpseudocode}
\usepackage{xcolor}
\usepackage{subfig}

\usepackage{url}

\graphicspath{{img/}}

\allowdisplaybreaks

\begin{document}


\title{AERIQ: SDR-Based LTE I/Q Measurement and Analysis Framework for Air-to-Ground Propagation Modeling}

\author{S. J. Maeng, O. Ozdemir, \.{I}. G\"{u}ven\c{c}, M. L. Sichitiu, R. Dutta, and M. Mushi\\
Department of Electrical and Computer Engineering, North Carolina State University, Raleigh, NC\\
{\tt \{smaeng,oozdemi,iguvenc,mlsichit,rdutta,mjmushi\}@ncsu.edu}}

\maketitle

\begin{abstract}
In this paper, we introduce AERIQ: a software-defined radio (SDR) based I/Q measurement and analysis framework for wireless signals for aerial experimentation. AERIQ is integrated into controllable aerial vehicles, it is flexible, repeatable, and provides raw I/Q samples for post-processing the data to extract various key parameters of interest (KPIs) over a 3D volume. Using SDRs, we collect I/Q data with unmanned aerial vehicles (UAVs) flying at various  altitudes in a radio dynamic zone (RDZ) like outdoor environment, from a 4G LTE eNB that we configure to operate at 3.51 GHz. Using the raw I/Q samples, and using Matlab's LTE Toolbox, we provide a step-by-step description for frequency offset estimation/correction, synchronization, cell search, channel estimation, and reference signal received power (RSRP). We provide various representative results for each step, such as RSRP measurements and corresponding analytical approximation at different UAV altitudes, coherence bandwidth and coherence time of the channel at different UAV altitudes and link distances, and kriging based 3D RSRP interpolation. 
The collected raw data as well as the software developed for obtaining and post-processing such data are provided publicly for potential use by other researchers. AERIQ is also available in emulation and testbed environments for external researchers to access and use as part of the NSF AERPAW platform at NC State University.
\end{abstract}


\section{Introduction}\label{sec:intro}
As the demand for advanced wireless communications rapidly rises, the need for supporting high transmission speeds, low latency, and massive connectivity is growing. 
To achieve these, efficient and flexible spectrum management systems that can help utilize unoccupied and under-utilized spectrum resources need to be developed and  validated. Recently, the concept of radio dynamic zones (RDZs) have been introduced, which are areas or  volumes with automatic spectrum management mechanisms that control electromagnetic energy entering, escaping, or occupying the zone~\cite{NSF_SIINRDZ_Solicitation,9814648}. RDZs can help develop, test, and improve novel spectrum sharing and coexistence technologies in the future to improve utilization of spectrum resources. 

In an RDZ, spectrum should be monitored and controlled in real-time to share the spectrum resources with incumbent users outside of the zone. 
This is challenging and complicated, as it is not feasible to deploy sensor nodes very densely at every location across a large RDZ area. However, spectrum measurements obtained at sparsely located sensor nodes, or over the trajectory of one or more autonomous unmanned aerial or ground vehicles (UAVs/UGVs) equipped with spectrum sensors, can be used to interpolate spectrum occupancy across an RDZ. For example, in~\cite{yilmaz2013radio}, radio environment maps (REMs) have been used for monitoring the coverage of cellular networks. In~\cite{sato2017kriging,7542153}, Kriging has been studied to spatially interpolate the signal power by using spectrum measurements from nearby locations. In~\cite{xue2014cooperative,jin2018privacy,thilina2013machine}, cooperative and crowd-sourced spectrum sensing have been investigated. In \cite{maeng2021out}, a signal leakage power monitoring technique is proposed for RDZs using sparsely deployed ground sensor nodes. 

\begin{figure}[t]
	\centering
	\includegraphics[width=0.48\textwidth]{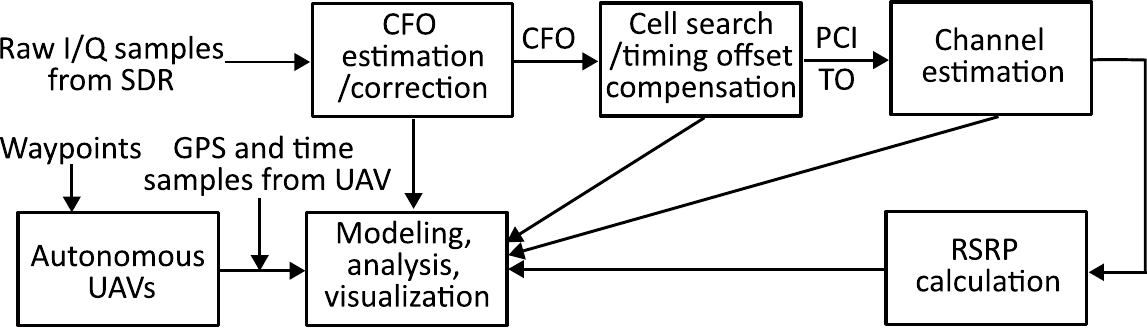}
	\caption{Block diagram for  post-processing of LTE I/Q samples collected at an SDR that is carried by a UAV.}\label{fig:block_diagram}
\end{figure}

In order to have real-time volumetric spectrum awareness in an RDZ, it is critical to understand the propagation characteristics throughout that RDZ, such as time, frequency, and spatial correlation  for various communications scenarios, transmitter/receiver configurations, and environmental conditions. In particular, an accurate understanding of the characteristics of air-to-ground channel propagation is critical to generate a real-time and volumetric radio map for spectrum monitoring. In the literature, \cite{8108204} measures and models air-to-ground path loss at an open rural area with 5 different heights from 20~m to 100~m using a cellular smartphone-mounted UAV, for which Qualipoc software from Rohde \& Schwarz is used to obtain reference signal received power (RSRP). In \cite{8369158}, LTE uplink measurements from a cellular network are recorded and analyzed  using a UAV carrying a smartphone with QualiPoc at two heights, 1.5~m, and 100~m. In \cite{8746290}, two different heights of smartphone-mounted UAVs are used to collect measurements from public LTE networks. Three different commercial measurement software, namely, Keysight NEMO Outdoor, Rohde \& Schwarz QualiPoc, and InfoVista TEMS are utilized to obtain four different key parameters of interest (KPIs):  physical resource blocks (PRBs) utilization, modulation and coding scheme (MCS) level, throughput, and transmit power. Data collected through such commercial smartphone software are limited to the set of KPIs that are supported by those software, and it is not possible to extract  additional information. Collecting raw I/Q samples in the desired band and frequency of interest gives the flexibility to extract KPIs that may not be available through the commercial software, and test new approaches for receiver processing.


In this paper, we propose AERIQ: a software-defined radio (SDR) based I/Q measurement and analysis framework with autonomous UAVs for aerial wireless experimentation. AERIQ  is flexible, repeatable, and provides raw I/Q samples for post-processing the data to extract various KPIs over a 3D volume of interest. We conduct  experiments at the NSF Aerial Experimentation and Research Platform for Advanced Wireless (AERPAW) Lake Wheeler Field Labs site~\cite{marojevic2020advanced} in an open rural area. Using SDRs, we collect I/Q data with UAVs flying at various different altitudes in an RDZ-like outdoor environment, from a 4G LTE eNB that we configure to operate at 3.51 GHz. Using the raw I/Q samples, and using Matlab's LTE Toolbox, we provide a step-by-step description for the following post-processing stages at an aerial receiver: frequency offset estimation/correction, synchronization, cell search, channel estimation, and RSRP calculation (see Fig.~\ref{fig:block_diagram}). 
We provide various representative results for each step, such as RSRP measurements and corresponding analytical approximation at different UAV altitudes, coherence bandwidth and coherence time of the channel at different UAV altitudes and link distances, and kriging based RSRP interpolation across a 3D volume.

The rest of this paper is organized as follows. In Section~\ref{sec:measurement_campaign}, we describe our measurement environment and setup, and elaborate on a UAV architecture and trajectory design. In Section~\ref{sec:CFO_est}, we present results on estimated carrier frequency offset (CFO) for different RSRP levels. In Section~\ref{sec:TO_Cellsearch}, timing offset (TO) detection and the cell search process are carried out using LTE synchronization sequences. In Section~\ref{sec:chan_est}, we show the estimated frequency channel response obtained by cell-specific reference signals (CRSs) in different heights and RSRPs. In Section~\ref{sec:RSRP}, we analyze and model RSRPs with different heights using two-ray path loss model and antenna radiation patterns. In Section~\ref{sec:coh_bw}, the coherence bandwidth and coherence time at different heights are analyzed, which are extracted from LTE channel estimation results. In Section~\ref{sec:spatial_cor}, we analyze and model shadowing components and spatial correlation, and finally,  Section~\ref{sec:conclusion} provides concluding remarks. 



\section{Measurement Campaign} \label{sec:measurement_campaign}

In the following subsection, we first briefly review the concept of RDZs and describe conducted experiments to measure and analyze the channel propagation between the transmitter on the ground and the aerial sensor node. The block diagram in Fig.~\ref{fig:block_diagram} describes how I/Q samples from measurements are processed in AERIQ. First, the CFO is estimated by a cyclic prefix (CP). Then, the timing offset and the physical cell identities (PCIs) are jointly estimated by the correlation between a received signal and candidate sequences of primary synchronization signal (PSS) and secondary synchronization signal (SSS). After the PCI is obtained, CRSs can be extracted and they are used in the channel estimation, RSRP calculation, and 3D RSRP interpolation stages. Note that the accuracy of the PCI detection, channel estimation, and hence, 3D RSRP interpolation depend heavily on the accuracy of time/frequency synchronization and cell search procedures at the receiver.

\begin{figure}[t]
	\centering
	\includegraphics[width=0.48\textwidth]{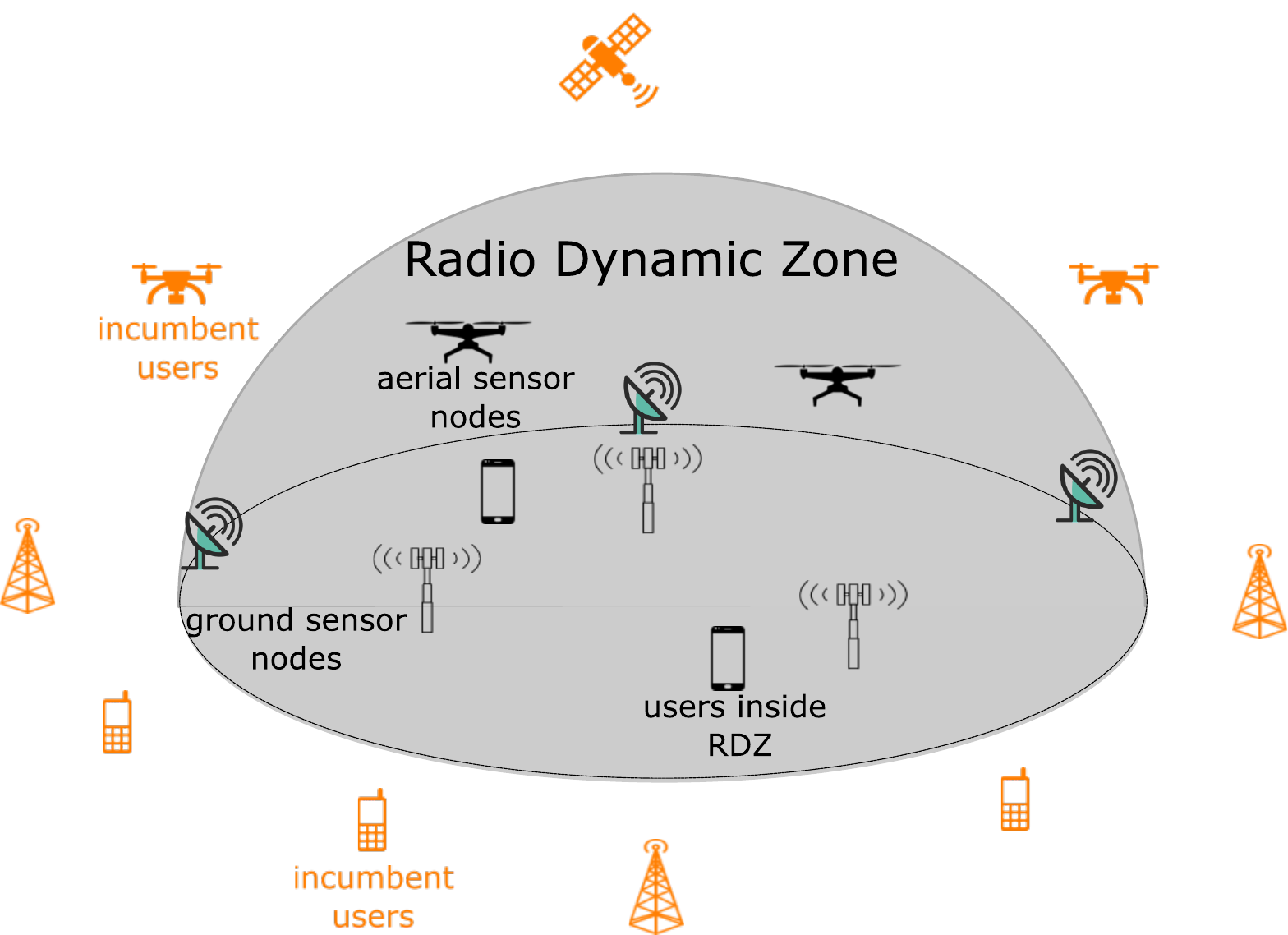}
	\caption{The illustration of an RDZ with aerial and ground sensors and users. The experiments carried out within an RDZ should not interfere with the users outside of the RDZ.}\label{fig:illu_RDZ}
\end{figure}

\subsection{Measurement Environment}

\begin{figure}[t]
	\centering
	\includegraphics[width=0.48\textwidth]{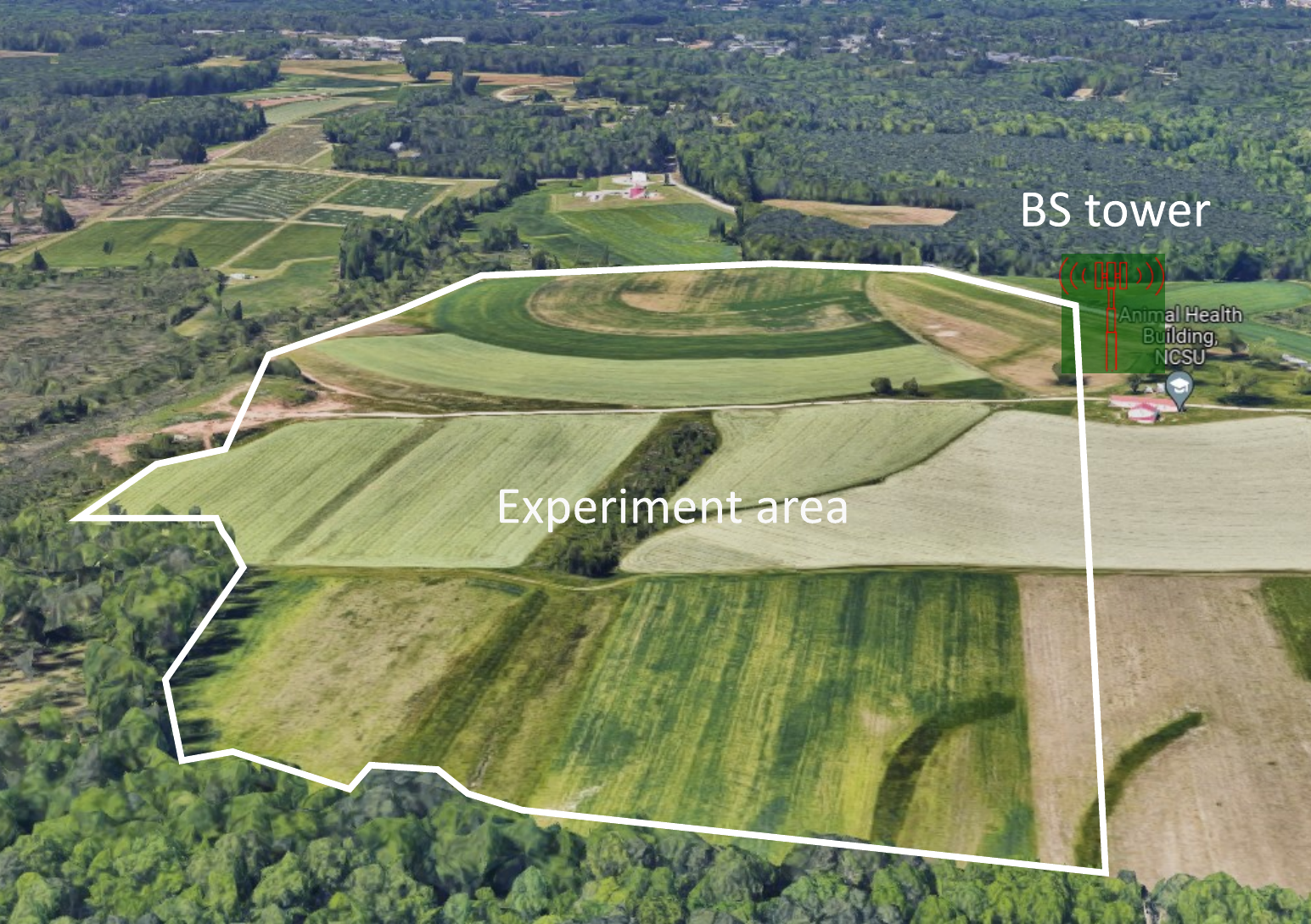}
	\caption{AERPAW's Lake Wheeler Field Labs site where air-to-ground I/Q samples have been collected.}\label{fig:exp_site}
\end{figure}

Before we describe our measurement setup, we will first overview our measurement environment. In this work, we consider an RDZ-like outdoor environment for advanced wireless and UAV experimentation as shown in Fig.~\ref{fig:illu_RDZ} and Fig.~\ref{fig:exp_site}. 
Spectrum sensing results across the RDZ at the frequencies of interest are assumed to be aggregated at a central processing unit, to determine whether the signal leakage from the zone results in interference problems with licensed receivers outside (and potentially, inside) the RDZ. For instance, if the signal leakage power level at the boundary of an RDZ exceeds a predefined threshold level, a transmission inside the zone can be automatically terminated. 
For continuous spectrum monitoring in the 3D space, ground and aerial spectrum sensor nodes can be deployed in an RDZ. In this paper, we consider measurements at one such monitoring node that is carried at a UAV at different altitudes.  

\subsection{Measurement Setup}

The measurement campaign is executed at the NSF AERPAW Lake Wheeler Road Field Labs site in Raleigh NC, USA, which is shown in Fig.~\ref{fig:exp_site}. A fixed BS tower with a single dipole transmit antenna is deployed, and the flying UAV is equipped with a vertically oriented single dipole receiver antenna and a GPS receiver. LTE eNB is realized using the srsRAN open-source SDR software at the BS tower that transmits the LTE downlink signal. An SDR is mounted on the UAV to collect raw I/Q data samples. For  both the tower and the UAV, a USRP B205mini from NI is used as an SDR. The UAV collects 20~ms segments of data out of every 100 ms, and out of that recording, and extracts a full LTE frame that is of length 10~ms. In other words, in every 100 ms, we capture one full LTE frame. The detailed specifications of the BS tower and the UAV receiver are listed in Table~I.

\subsection{UAV Architecture and Trajectory}
\begin{figure}[t]
	\centering
	\includegraphics[width=0.3\textwidth]{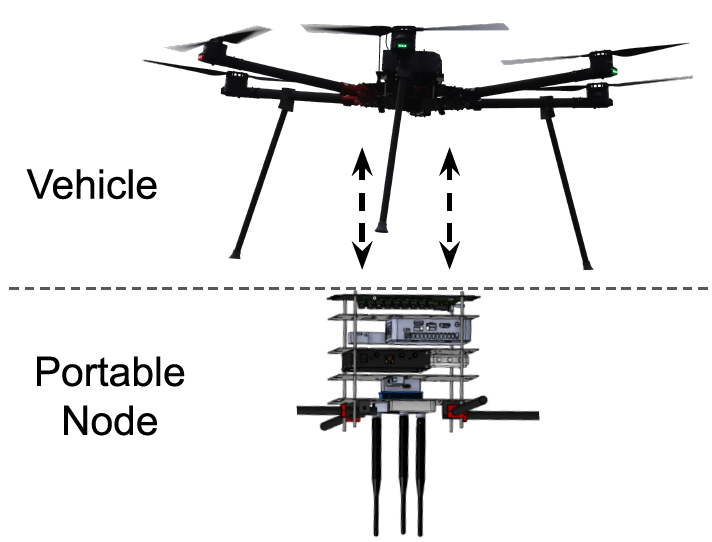}
	\caption{Illustration of the AERPAW UAV  that carries a portable node including an SDR~\cite{funderburk2022aerpaw}.}\label{fig:drone_architecture}
\end{figure}
\begin{figure}[t]
	\centering
	\includegraphics[width=0.48\textwidth]{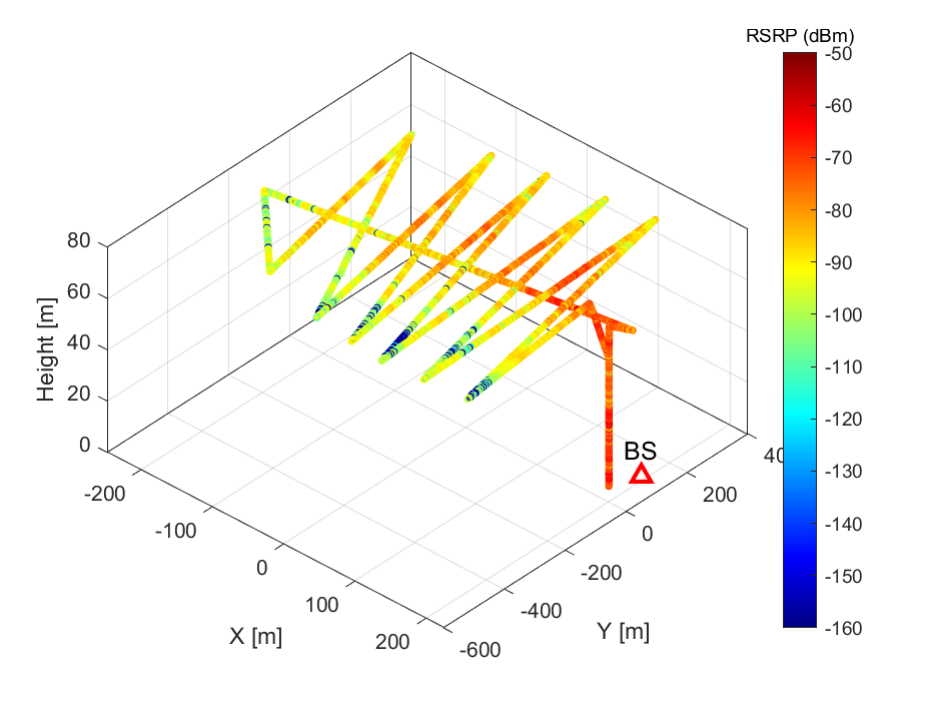}
	\caption{Trajectory and RSRP at UAV altitude of  70 m.}\label{fig:trajectory}
\end{figure}

In the experiments, we use a custom built UAV that carries a portable node as shown in Fig.~\ref{fig:drone_architecture}~\cite{funderburk2022aerpaw}. The portable node includes an NI USRP B205mini SDR that executes a Python script~\cite{AERPAW_site} to collect I/Q samples at the desired center frequency with the desired sampling rate.  
The UAV that carries the portable node flies on a predesigned trajectory at a fixed height. In particular, the UAV moves on a \emph{zig-zag} path through the experiment site and flies back to the starting position. The waypoints for the UAV are pre-programmed by the experimenter in a development environment and tested using emulations prior to the actual experiment for intended operation. Autonomous navigation based on real-time RF signal observations are possible (e.g., to improve spectrums sensing performance based on real-time trajectory decisions), but not considered in this paper.  The flights are conducted multiple times by increasing the height of the UAV (30~m, 50~m, 70~m, 90~m, 110~m), following an identical horizontal trajectory at each altitude. As an example, the UAV trajectory at an altitude of 70~m as well as the RSRP observed at each UAV location is shown in Fig.~\ref{fig:trajectory}.

\begin{table}[t]
\renewcommand{\arraystretch}{1.1}
\caption{Measurement setup for experiments.}
\label{table:settings}
\centering
\begin{tabular}{lc}
\hline
\textbf{BS Tower (Transmitter)} \\
\hline
Technology & LTE\\
Tower height & 10~m\\
Transmit power & 10~dBm\\
Carrier frequency & 3.51~GHz\\
Bandwidth & 1.4~MHz\\
Antenna & Dipole antenna (RM-WB1)\\
\hline
\hline
\textbf{UAV (Receiver)} \\
\hline
Antenna & Dipole antenna (SA-1400-5900)\\
UAV heights & \{30,~50,~70,~90,~110\}~m\\
\hline
\end{tabular}
\vspace{-0.15in}
\end{table}

\subsection{Altitude and Speed of the UAV by GPS Reading}

\begin{figure}[t]
	\centering
	\includegraphics[width=0.48\textwidth]{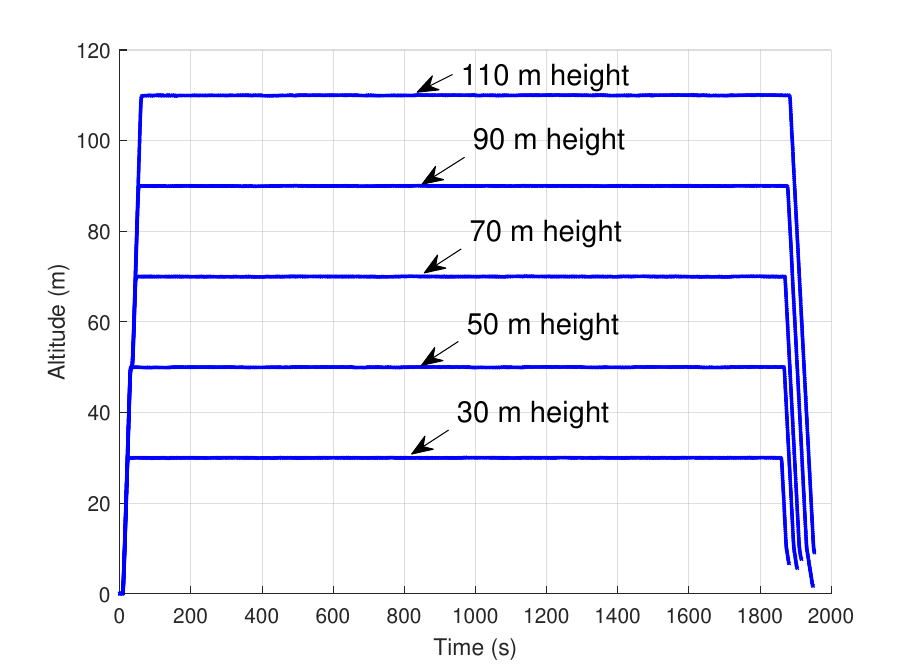}
	\caption{Altitude of the UAV from GPS logs versus time  for the five different experiments.}\label{fig:altitude}
\end{figure}
\begin{figure}[!t]
	\centering
	\includegraphics[width=0.48\textwidth]{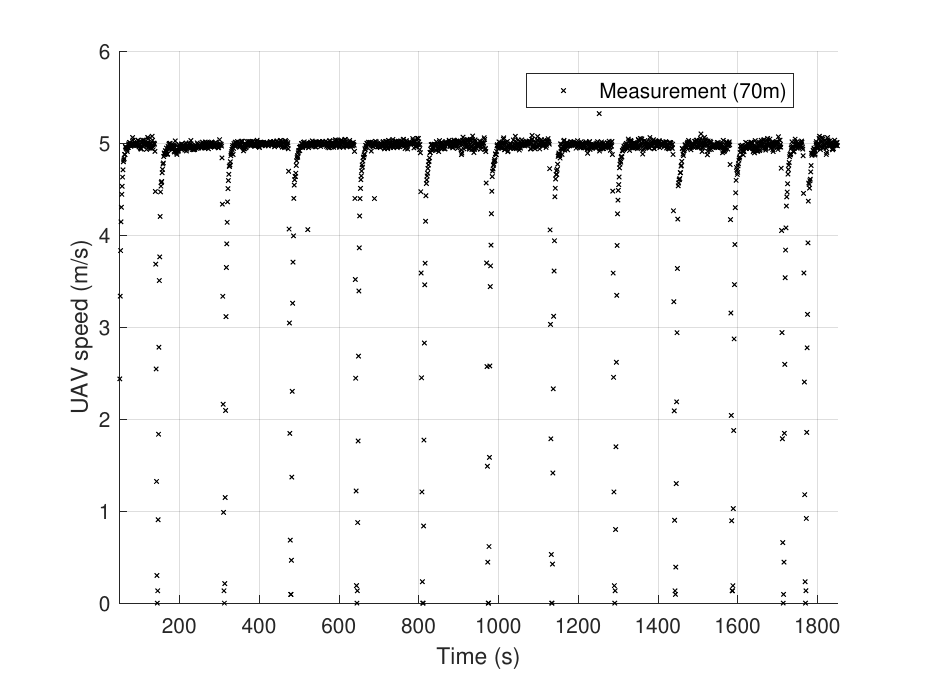}
	\caption{UAV speed versus time from GPS logs at UAV altitude 70~m.}\label{fig:UAV_speed}
\end{figure}

During the flight, the UAV flies up to the predefined altitude and flies horizontally at that altitude until landing. Fig.~\ref{fig:altitude} shows the change of the UAV altitude versus time for experiments at  different altitudes, which is obtained from the log of the GPS mounted on the UAV. Fig.~\ref{fig:UAV_speed} shows the UAV speed during the flight. The UAV speed can be obtained by the GPS locations and timestamps. It is shown that the UAV flies with constant speed and the UAV quickly slows down the speed when it changes direction at a certain waypoint, and it quickly recovers the speed right after that.

The GPS logs give information on the timestamp, the latitude, the longitude, and the altitude during flights, which is used to generate the UAV trajectory in Fig.~\ref{fig:trajectory} as well.


After we collect raw I/Q samples from the experiments, we process data to analyze the characteristic of the channel propagation by using MATLAB LTE Toolbox~\cite{LTE_Toolbox,vasudeva2020vehicular}. The post-processing includes cell ID search,  synchronization, and channel estimation as well as obtaining the RSRP.

\section{Carrier Frequency Offset Estimation}\label{sec:CFO_est}
\begin{figure}[t]
	\centering
	\subfloat[70 m height at 380 s (70~m, RSRP high).]{\includegraphics[width=0.48\textwidth]{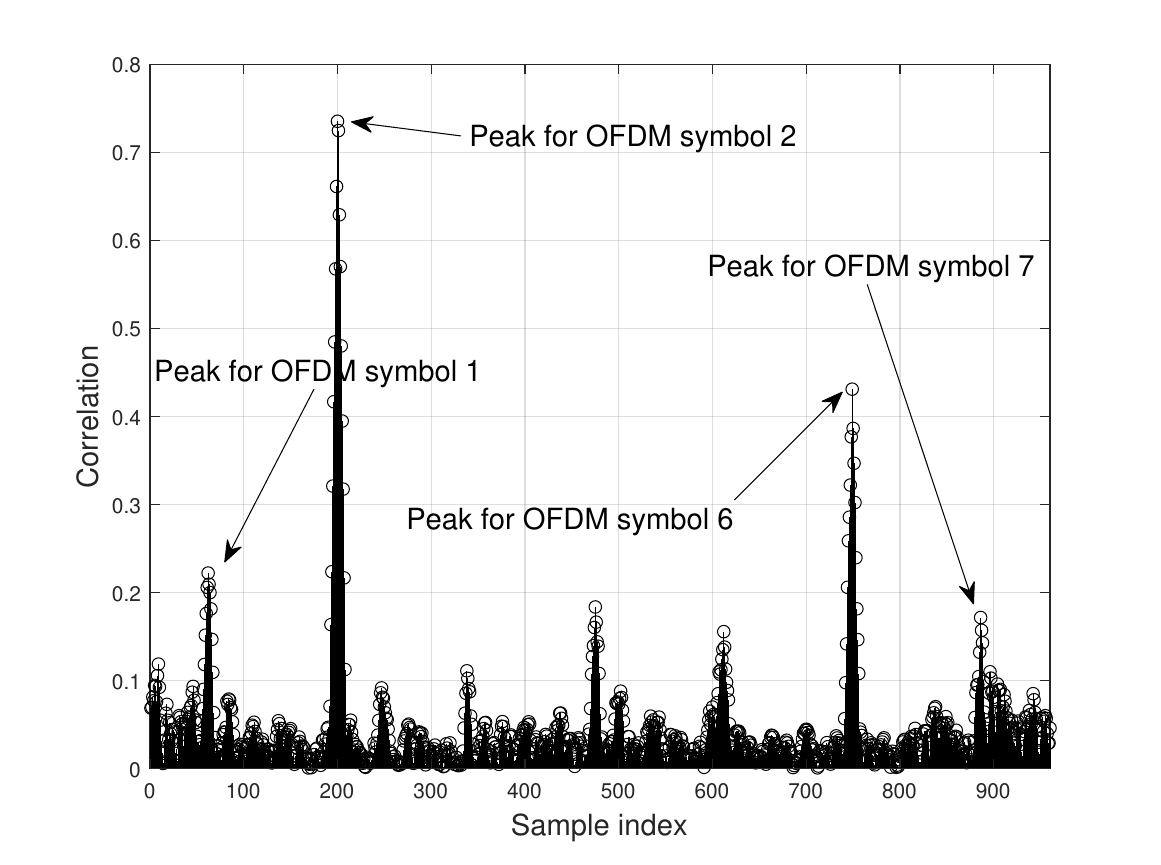}}
 
        \subfloat[70 m height at 520 s (70~m, RSRP low).]{\includegraphics[width=0.48\textwidth]{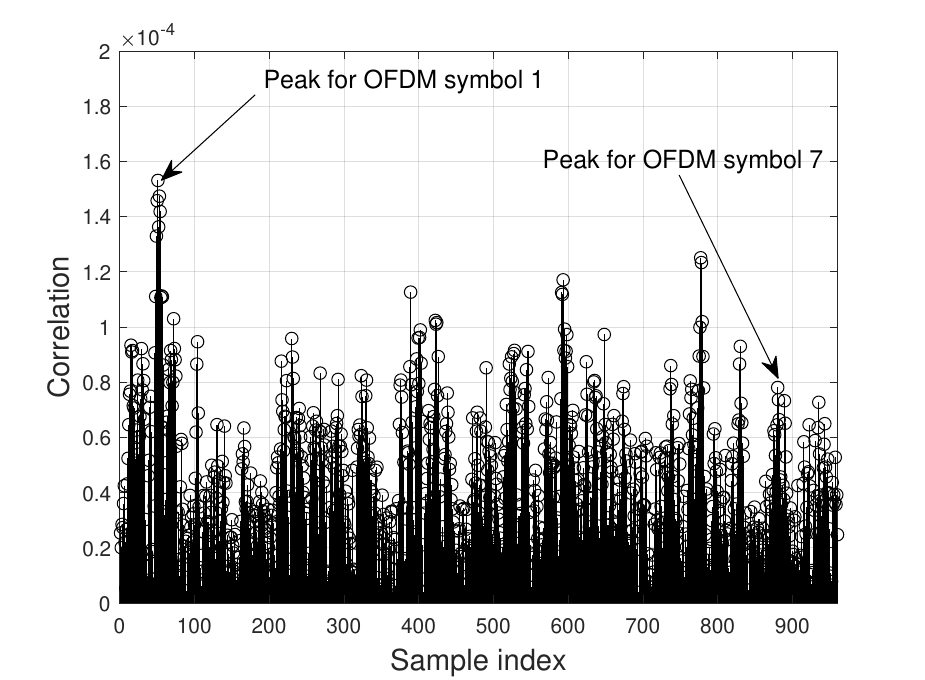}}
	\caption{Moving correlation with CP length window for 1 slot length (7 OFDM symbols, 0.5 ms). The peak position indicates the starting of the CP.}\label{fig:CFO_corr}
\end{figure}
\begin{figure}[t]
	\centering
	\includegraphics[width=0.48\textwidth]{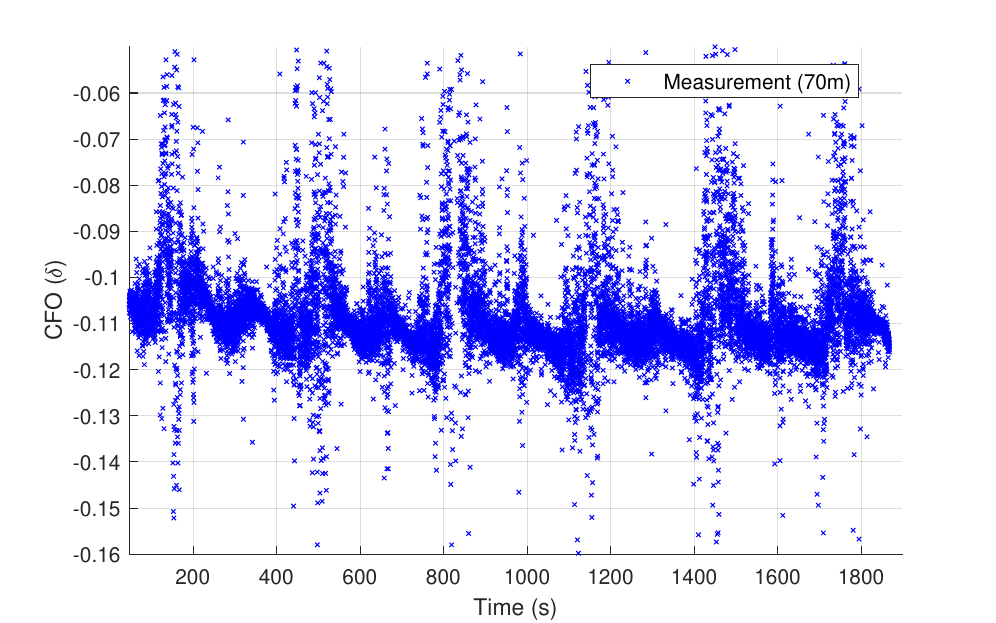}
	\caption{Estimated CFO (fracton of subcarrier spacing) versus time at UAV altitude of 70~m.}\label{fig:CFO_est}
\end{figure}

The CFO can occur from the instability of the oscillators between Tx and Rx, or Doppler shift(s) that occur due to the velocity difference between a transmitter and a receiver. In an LTE system, the CFO can be estimated in the time-domain waveform by using the CP, which is the repeated version of the fraction of the orthogonal frequency-division multiplexing (OFDM) symbols attached in front of the OFDM symbols. The time-domain OFDM signal of the CP with CFO can be represented as follows:
\begin{align}
    y[n]&=x[n]e^{j2\pi n\delta/N},
\end{align}
where $\delta$, $N$ denotes CFO and the number of subcarriers. Note that the unit for the CFO $\delta$ is subcarrier spacing. The repeated OFDM signal corresponding to the CP is given by
\begin{align}
    y[n+N]&=x[n+N]e^{j2\pi (n+N)\delta/N}=x[n]e^{j2\pi\delta + j2\pi n\delta/N}.
\end{align}
The CFO can be obtained by the correlation between CP and the corresponding OFDM symbol, as
\begin{align}\label{eq:cfo_est}
    \hat{\delta}&=\frac{1}{2\pi}\arg\left(\frac{1}{{\rm L}_{\rm cp}}\sum_{n=0}^{{\rm L}_{\rm cp}-1}x[n]^{\star}x[n+N]\right),
\end{align}
where ${\rm L}_{\rm cp}$ is the length of the CP and $(\cdot)^{\star}$ indicates the complex conjugate operation. 

We use collected LTE I/Q samples from experiments to evaluate the reliability of the CFO estimation. We consider the I/Q  measurements at the UAV altitude of 70~m for evaluation. Fig.~\ref{fig:CFO_corr} shows the $N$-shifted moving auto-correlation of the time-domain received signal with the ${\rm L}_{\rm cp}$ window size for 1 slot length (0.5~ms) to estimate the CFO $\delta$. The peak position indicates the starting of the CP. We show the correlations at two different UAV locations, where the  RSRP is high and low, as marked in Fig.~\ref{fig:RSRP_time_all}. It is observed that the correlation peaks for each OFDM symbol are clearly visible at high RSRP, while they are not as easy to distinguish at low RSRP.

Fig.~\ref{fig:CFO_est} shows the estimated CFO versus time using \eqref{eq:cfo_est} at the UAV altitude of 70~m in Fig.~\ref{fig:trajectory} and Fig.~\ref{fig:altitude}. It is observed that the estimated CFO ($\hat{\delta}$) fluctuates between -0.13 and -0.09 during a whole flight, which shows that the trend of the estimated CFO is matched with the constant UAV speed as well as the slowing/accelerating UAV instants in Fig.~\ref{fig:UAV_speed}, while high sparks are also observed many times, which results from the estimation error at low RSRP. The estimated CFO is compensated before the timing synchronization and cell search stage which is dicusssed in the next section.

\section{Timing Offset Detection and Cell Search}\label{sec:TO_Cellsearch}
\begin{figure}[t!]
	\centering
	\subfloat[Correlation by PSS (RSRP high, 70~m, 380 s).]{\includegraphics[width=0.44\textwidth]{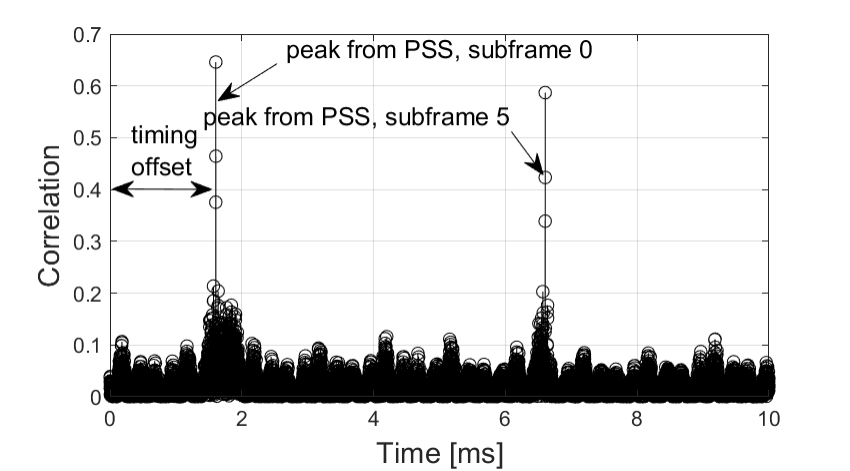}}

        \subfloat[Correlation with both  PSS, SSS subframe 0, 5 (RSRP high, 70~m, 380 s).]{\includegraphics[width=0.44\textwidth]{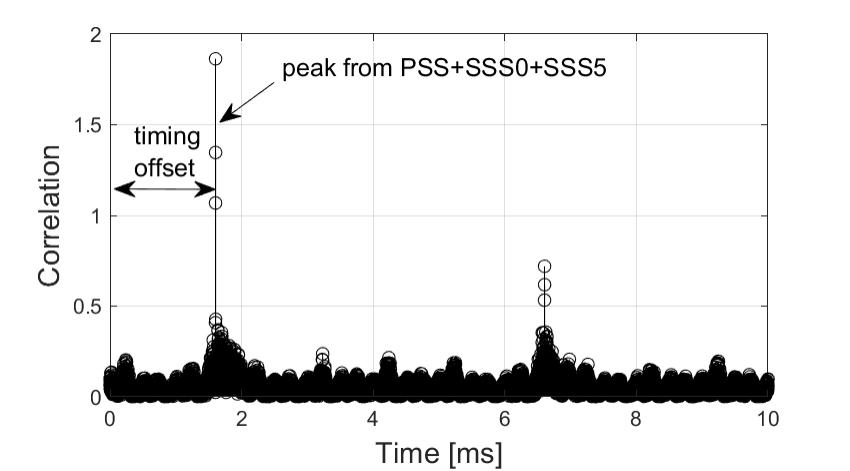}}
        
        \subfloat[Correlation by PSS (RSRP low, 70~m, 520 s).]{\includegraphics[width=0.44\textwidth]{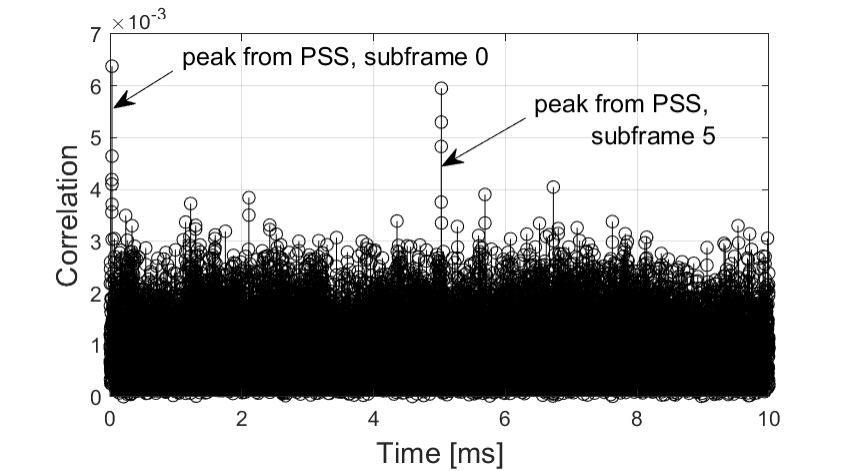}}
        
        \subfloat[Correlation with both PSS and SSS subframe 0, 5 (RSRP low, 70~m, 520 s).]{\includegraphics[width=0.44\textwidth]{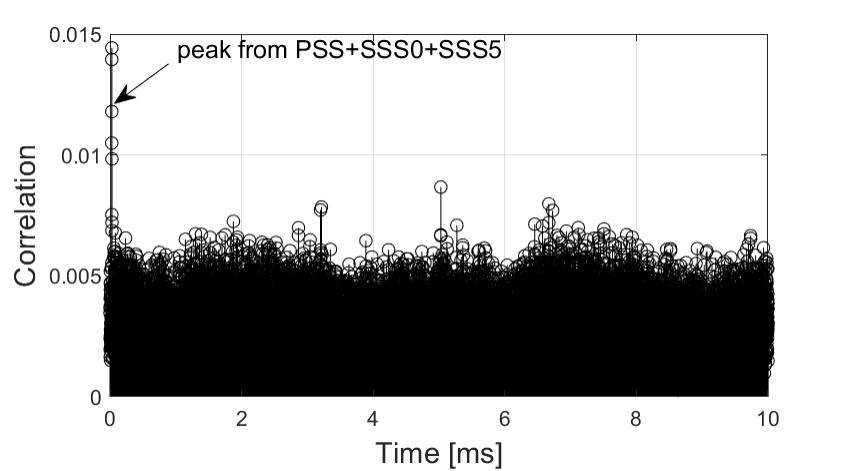}}
	\caption{The correlation of synchronization sequences to detect the timing offset at high RSRP ((a) and (b)) and low RSRP ((c) and (d)) locations.}\label{fig:sync_corr}
\end{figure}

The time offset from the received signal can be detected by using LTE synchronization sequences:  PSS and SSS. The PSS in LTE is generated from Zadoff–Chu sequences and the SSS is designed using M-sequences, both of which have good auto-correlation properties~\cite{sesia2011lte}. The correlation peak location indicates the position of the received PSS and SSS, both of which are always carried at the first slot of subframes 0 and 5 within a frame for FDD-based LTE transmission. Timing offset can be found by the correlation of PSS as well as the combination of the correlation of the PSS, the SSS subframe 0, and subframe 5 to improve the accuracy of the detection. More specifically, individual correlations of the incoming received signal with the PSS and the SSS subframe 0, and subframe 5 can be added to improve the detection performance. The correlation of the SSS subframe 5 is cyclic-shifted to align with the SSS subframe 0 correlation when three different correlations are summed. The timing offset detection from synchronization signals can be expressed as
\begin{align}
    \hat{m}&=\arg\max_{m}\left(\frac{1}{{\rm L}_{\rm ss}}\sum_{n=0}^{{\rm L}_{\rm ss}-1}y[n+m]s[n]^{\star}\right),
\end{align}
where $s[n]$ and ${\rm L}_{\rm ss}$ denote the synchronization signal and the length of the synchronization signal, respectively. Note that while we consider non-coherent detection for both the PSS and the SSS in this work, the SSS can be  detected by coherent detection as also discussed in~\cite[Ch.~7]{sesia2011lte}.

The PSS and SSS distinguish 504 unique PCIs, which can be estimated by the combination of two different ID groups from PSS and SSS. The physical layer ID 0 - 2 is given by the PSS, and another physical layer group ID 0 - 167 is provided by the SSS. Among all possible physical layer ID cases, the physical layer ID that achieves the highest correlation peak of the synchronization signals is chosen as the correct physical layer ID. The PCI (${\rm N}_{\rm pci}$) is estimated by
\begin{align}
    \hat{{\rm N}}_{\rm pci}&= \hat{{\rm N}}_{\rm ID}^{\rm pss} + 3\hat{{\rm N}}_{\rm ID}^{\rm sss},\label{Eq:PCI}
\end{align}
where $\hat{{\rm N}}_{\rm ID}^{\rm pss}$, $\hat{{\rm N}}_{\rm ID}^{\rm sss}$ indicate the estimated physical layer ID and group ID, from the PSS and the SSS, respectively.

Fig.~\ref{fig:sync_corr} shows the correlation of the synchronization sequences to detect the timing offset. We observe correlations at high and low RSRP at 70~m height measurement. We can observe that the peak is relatively high and clear when we find the timing offset combing PSS and SSSs. Furthermore, it is observed that the peaks are detected at low RSRP as well, which implies that the timing offset is properly detected at low RSRP.

\section{Channel Estimation}\label{sec:chan_est}

\begin{figure}[t]
	\hspace{-2mm}\subfloat[70~m height at 380~s (RSRP high)]{\includegraphics[width=0.54\textwidth]{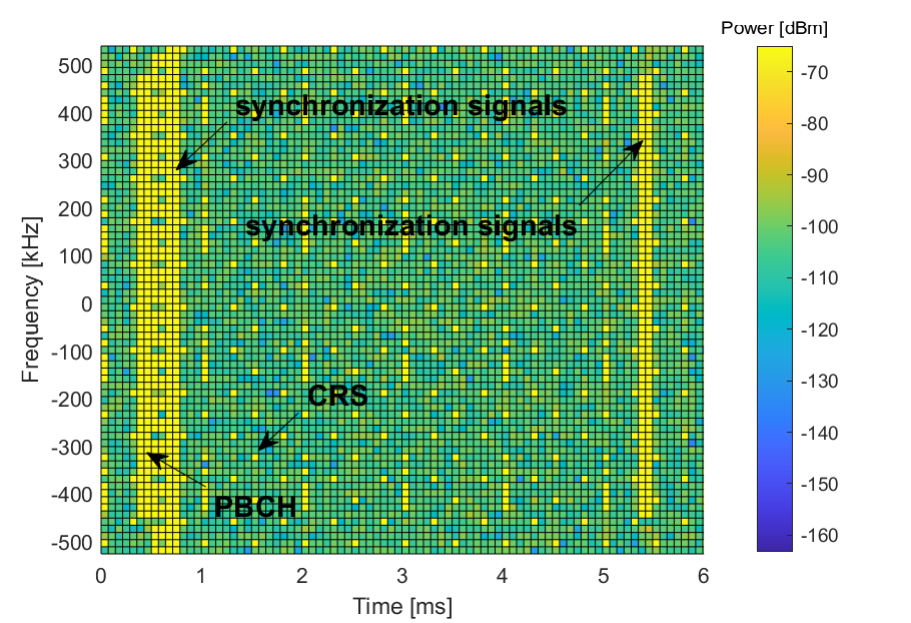}}

        \hspace{-2mm}\subfloat[70~m height at 520~s (RSRP low)]{\includegraphics[width=0.54\textwidth]{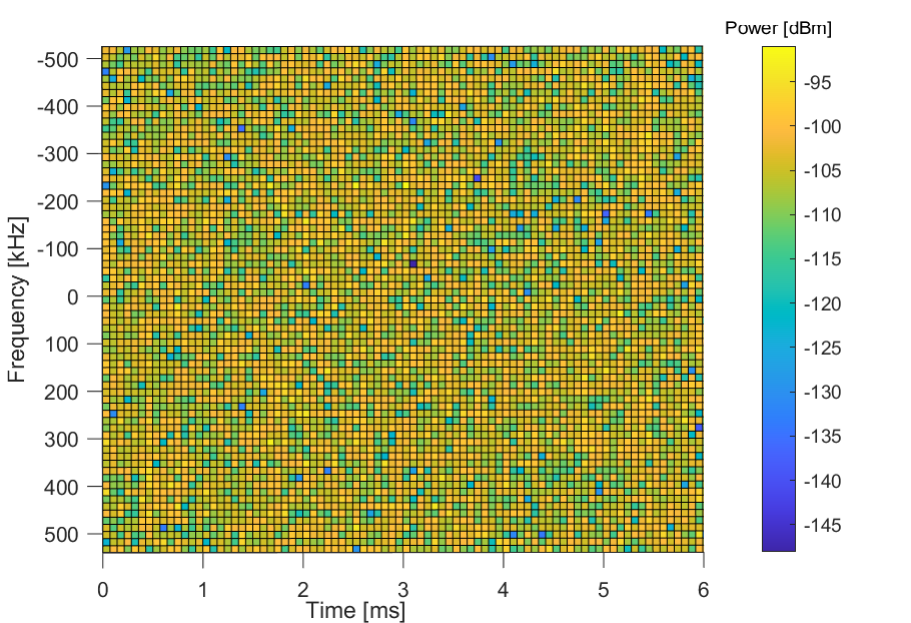}}
	\caption{Received signal power in LTE resource grid of duration six subframes and having a bandwidth of 1.08~MHz.}\label{fig:res_grid}
\end{figure}


\begin{figure}[t!]
	\centering
	\subfloat[70~m height at 380~s (RSRP high)]{\includegraphics[width=0.44\textwidth]{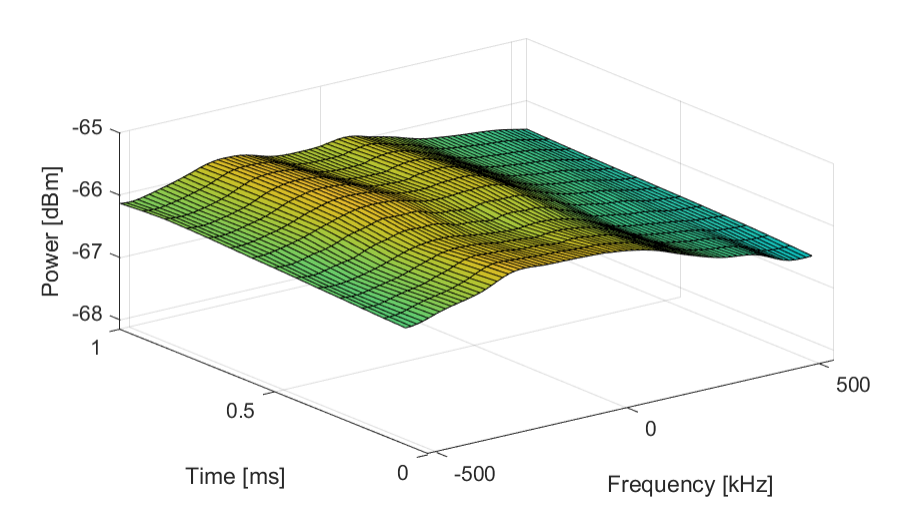}\label{fig:chan_est_1}}

        \subfloat[110~m height at  380~s (RSRP high)]{\includegraphics[width=0.44\textwidth]{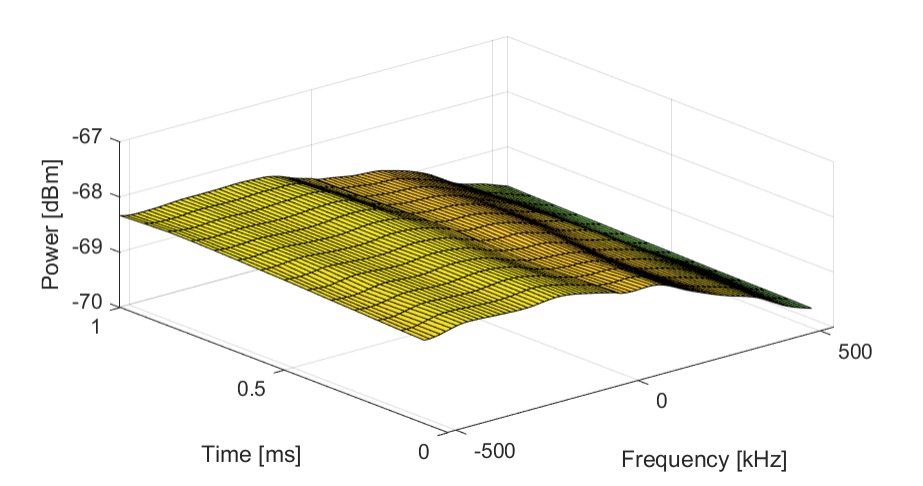}\label{fig:chan_est_2}}

        \subfloat[70~m height at 520~s (RSRP low)]{\includegraphics[width=0.44\textwidth]{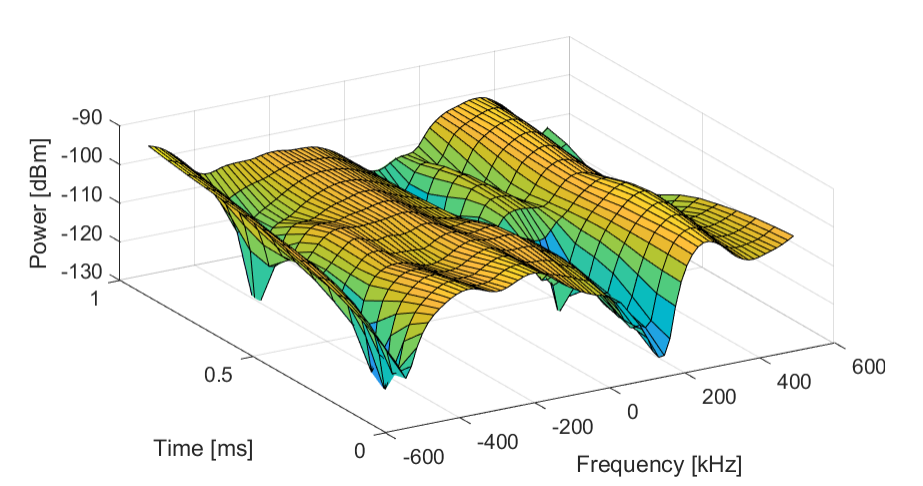}\label{fig:chan_est_3}}

        \subfloat[110~m height at 520~s (RSRP low)]{\includegraphics[width=0.44\textwidth]{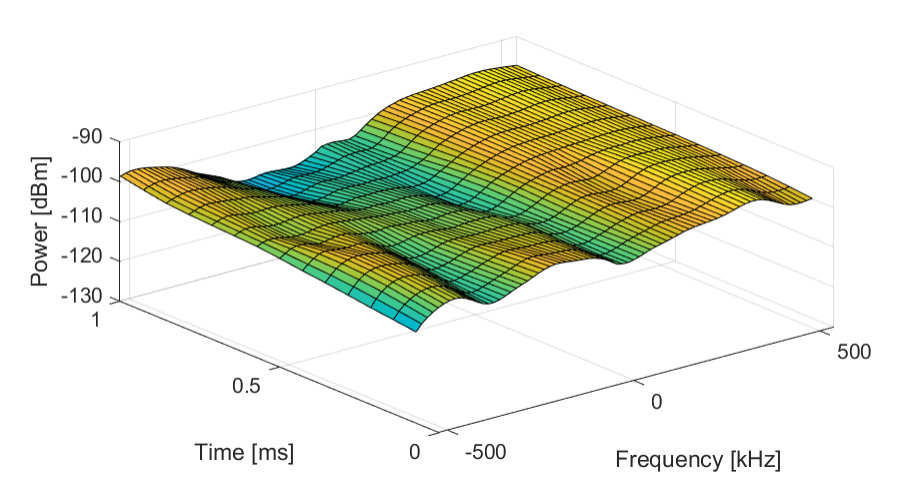}\label{fig:chan_est_4}}
	\caption{The estimated channel of 70~m and 110~m height flights at the time that RSRP is low and high.}\label{fig:chan_est}
\end{figure}

After the synchronization and the cell search procedures end, the LTE CRS, which is sparsely allocated in the resource grid, can be extracted in the frequency domain to estimate the channel. The CRS transmitted by a certain eNB is uniquely connected to the PCI of that eNB, calculated in~\eqref{Eq:PCI}. In particular, the CRS location is shifted by $\hat{{\rm N}}_{\rm pci}~\textrm{mod}~6$ in the frequency domain. Moreover, the CRS  for a given cell is generated from a Gold sequence that is uniquely characterized by the PCI of that cell.  

The channel coefficients of the CRS locations can be estimated by the least square (LS) method and then channel coefficients of the unknown locations can be interpolated by cubic interpolation~\cite[Ch.~8]{sesia2011lte}. Fig.~\ref{fig:res_grid} shows the received signal power in the frequency domain over the LTE resource grid that has a bandwidth of $1.08$~MHz and is observed over $6$~subframes. It can be seen that the physical broadcast channel (PBCH) is transmitted at subframe 0, the synchronization signals are allocated at subframe 0 and 5, and the CRSs are continuously transmitted across the whole grid with a specific pattern. Note that we set up the experiments so that other LTE data and control channels are not transmitted during the UAV flight, which can be observed by the unoccupied resources in the received signal in Fig.~\ref{fig:res_grid}. 

The estimated time and frequency domain channel for UAV altitudes of 70~m and 110~m are shown in Fig.~\ref{fig:chan_est}. After the UAV reaches the desired height and the UAV is located close to the BS tower, the RSRP is high and the channel in the frequency domain varies only slightly for both altitudes of 70~m and 90~m as shown in Fig.~\ref{fig:chan_est_1} and Fig.~\ref{fig:chan_est_2}, respectively. However, when the UAV is far away from the BS tower in Fig.~\ref{fig:chan_est_3} and Fig.~\ref{fig:chan_est_4},  RSRP is low and the channel in both time and frequency domains fluctuate, which is more visible for the 70~m UAV height when compared with the 90~m UAV height.


\section{RSRP Calculation from CRS}\label{sec:RSRP}
\begin{figure}[t]
	\centering
	\includegraphics[width=0.48\textwidth]{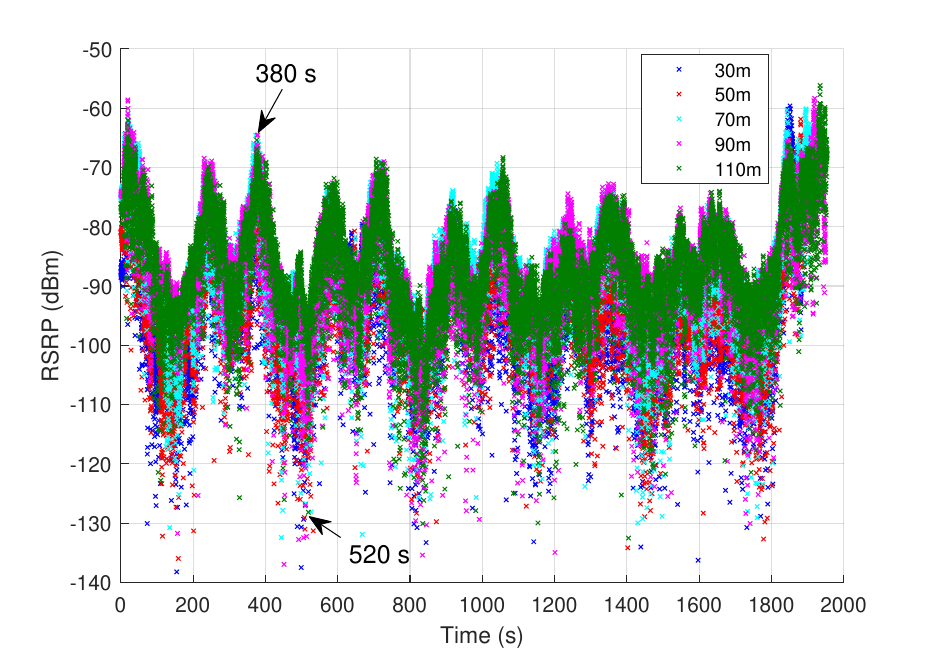}
	\caption{Change of the RSRP over time at different altitudes.}\label{fig:RSRP_time_all}
\end{figure}

\begin{figure}[t!]
	\centering
	\subfloat[Tx antenna (2.6 GHz)]{\includegraphics[width=0.24\textwidth]{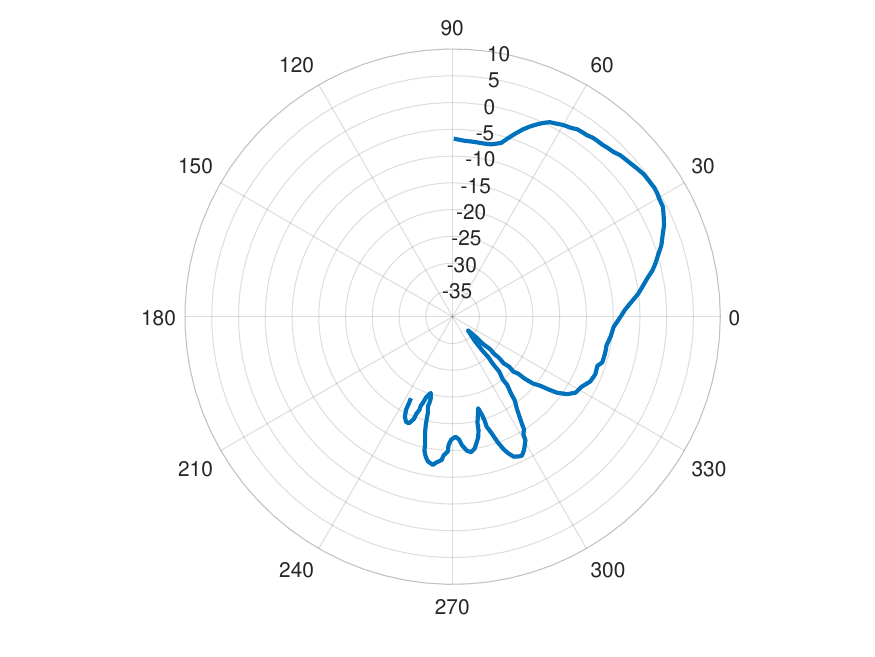}}~
        \subfloat[Rx antenna (2.4 GHz)]{\includegraphics[width=0.24\textwidth]{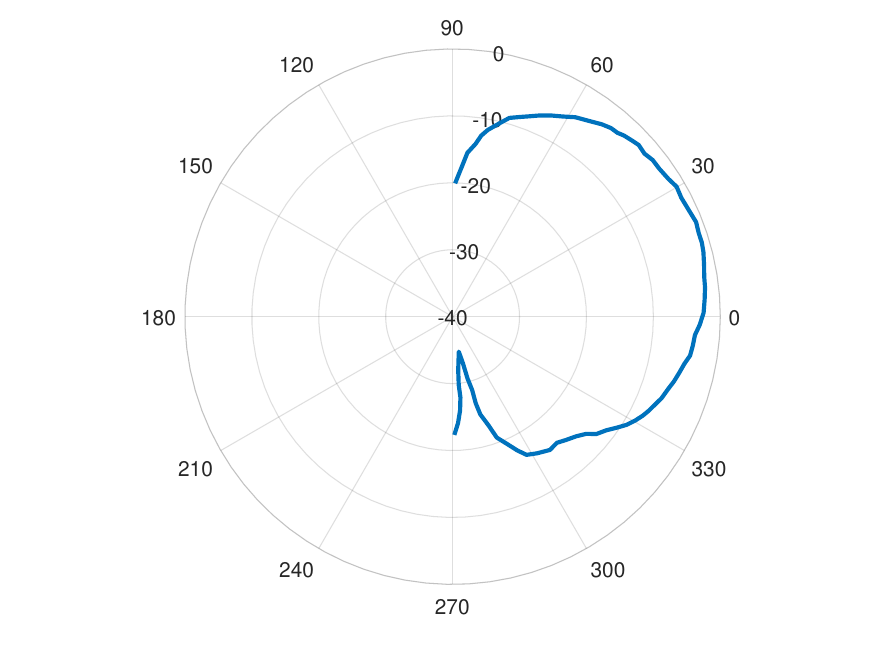}}
	\caption{The dipole antenna radiation patterns.}\label{fig:ant_pat}
\end{figure}

\begin{figure}[t!]
	\centering
	\subfloat[30~m height]{\includegraphics[width=0.48\textwidth]{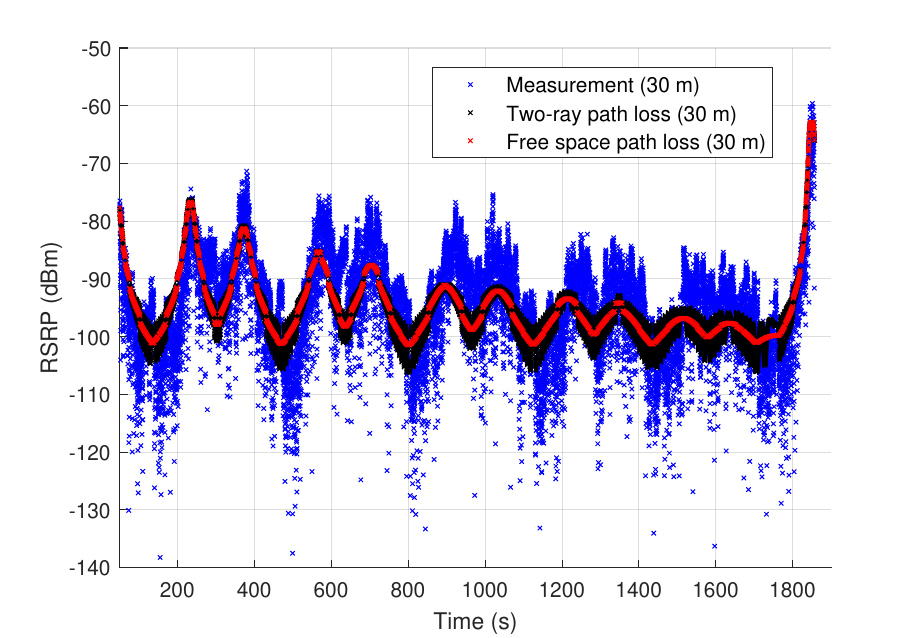}}
 
        \subfloat[70~m height]{\includegraphics[width=0.48\textwidth]{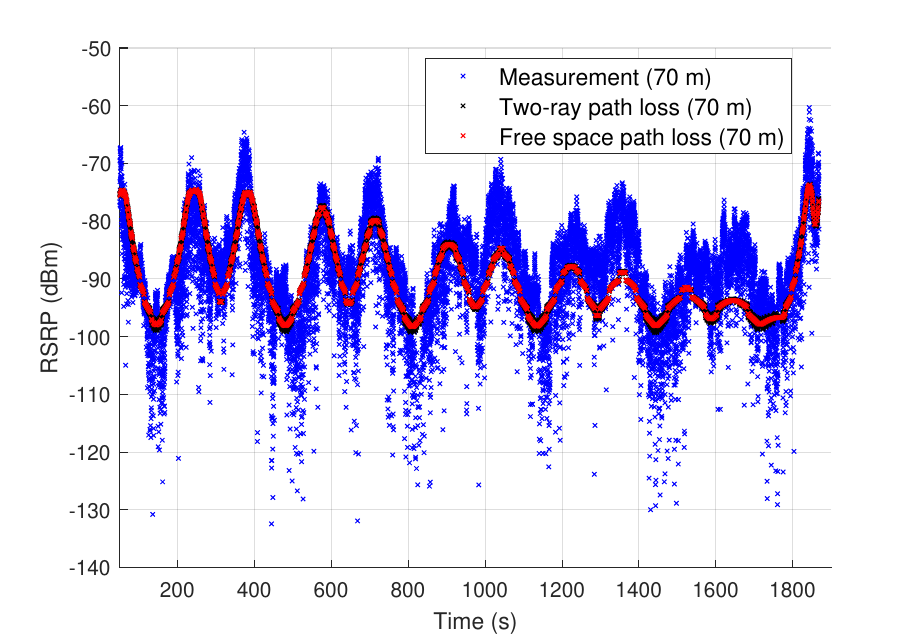}}
        
        \subfloat[90~m height]{\includegraphics[width=0.48\textwidth]{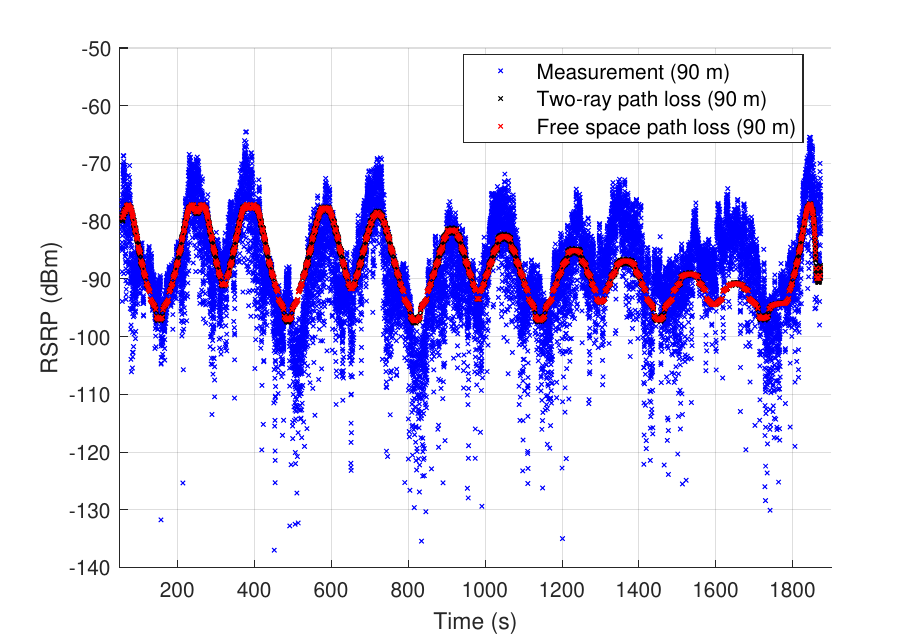}}
	\caption{The RSRP fitting using free-space and two-ray path loss models.}\label{fig:RSRP_PL}
\end{figure}

RSRP indicates the average signal strength that is calculated over the CRS and it averages out the effects of interference.  Fig.~\ref{fig:RSRP_time_all} shows the RSRP obtained during the experiments at different UAV altitudes. It is observed that the RSRP overall behavior is similar at all heights, which results from using identical horizontal trajectories for different flights. Because of the pattern of the UAV trajectory, the RSRP increases and decreases multiple times during the flights, based on  the distance between the Tx tower and the UAV. We consider two extremes, 380 seconds and 520 seconds as marked on Fig.~\ref{fig:RSRP_time_all}, to correspond to high-RSRP and low-RSRP locations, respectively. Representative results on CFO estimation, timing offset estimation, received power over the resource grid, and channel estimation have been provided at these two locations earlier in this paper. 

We also analyze the measured RSRP by fitting it to different path loss models. In particular, we consider the free-space path loss model and the two-ray path loss model with  3D antenna radiation pattern for the Tx and the Rx. The antenna patterns at the elevation angle domain, which are provided by the antennas specification sheets~\cite{rm_wb1,sa_1400}, are shown in Fig.~\ref{fig:ant_pat}. Since the antenna specification sheets do not provide the data at 3.5~GHz carrier frequency that we carried the experiments at, we used radiation patterns from other carrier frequencies available in the specifications.  As we use dipole antennas for both the Tx and the Rx, the omni-directional antenna radiation can be assumed at the azimuth plane. 

The two-ray path loss model considers the line-of-sight and the strong ground reflected paths which are combined at the received signal. In particular, the two-ray path loss model is given as
\begin{align}\label{eq:PL_two}
    &\mathsf{PL}=\left(\frac{\lambda}{4\pi}\right)^2\nonumber\\
    &\times\left|\frac{\sqrt{\mathsf{G}_{\rm bs}(\theta_l)\mathsf{G}_{\rm uav}(\theta_l)}}{d_{\rm 3D}}+\frac{\Gamma(\theta_r)\sqrt{\mathsf{G}_{\rm bs}(\theta_r)\mathsf{G}_{\rm uav}(\theta_r)}e^{-j\Delta\tau}}{r_1+r_2}\right|^2,
\end{align}
where $\mathsf{G}_{\rm bs}(\theta)$, $\mathsf{G}_{\rm uav}(\theta)$, $\lambda$ denote the Tx antenna gain, the Rx antenna gain, and the wave-length, respectively, $\theta_r$ represents ground reflection angle, $\Delta\tau$ indicates the phase difference between two paths, and $\Gamma(\theta_r)$ denotes the ground reflection coefficient that is dependent on the material on the ground. In addition, $d_{\rm 3D}$, $r_1$, $r_2$ respectively represent the LoS distance from Tx to Rx, the distance from Tx to the ground of the ground reflection path, and the distance from the ground to the Rx of the ground reflection path. The fitted curves from the different heights of RSRP measurements are shown in Fig.~\ref{fig:RSRP_PL}. It is observed that the fitted path loss curves follow the measured RSRP for all heights. We also observe that the RSRP fluctuation from the ground reflected signal in the two-ray path loss model  is captured clearly in 30~m height, while the fluctuation is small in 70~m and 90~m heights. This is because higher height reduces Tx and Rx antenna gains of the ground reflection path. Similar modeling and analysis can also be carried out to characterize the reference symbol received quality (RSRQ) measurements in LTE.

\section{Analysis of Coherence Bandwidth and Coherence Time}\label{sec:coh_bw}
\begin{figure}[t]
	\centering
	\includegraphics[width=0.48\textwidth]{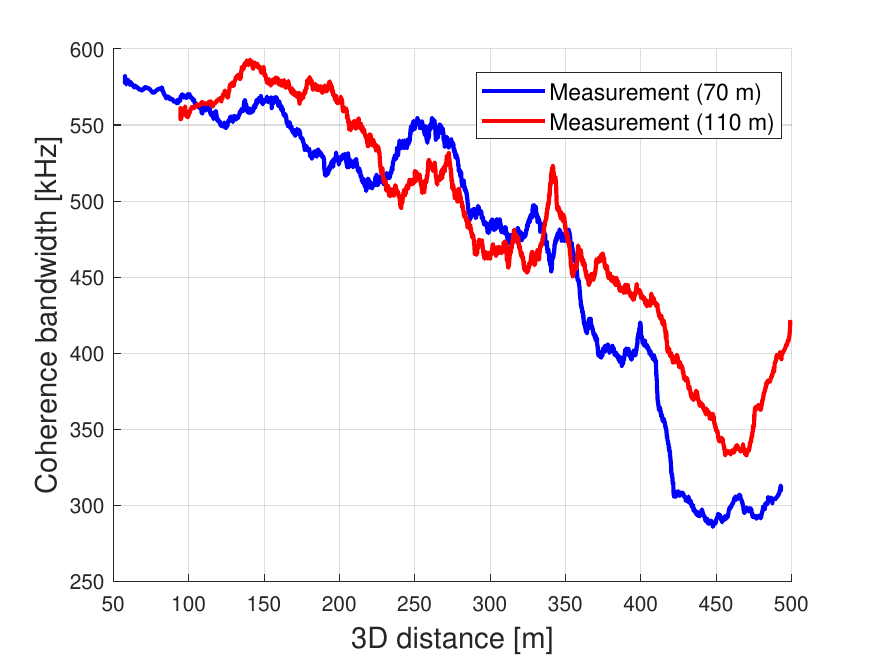}
	\caption{The coherence bandwidth depends on the 3D distance and the flight height.}\label{fig:coh_bw}
\end{figure}
\begin{figure}[t]
	\centering
	\includegraphics[width=0.48\textwidth]{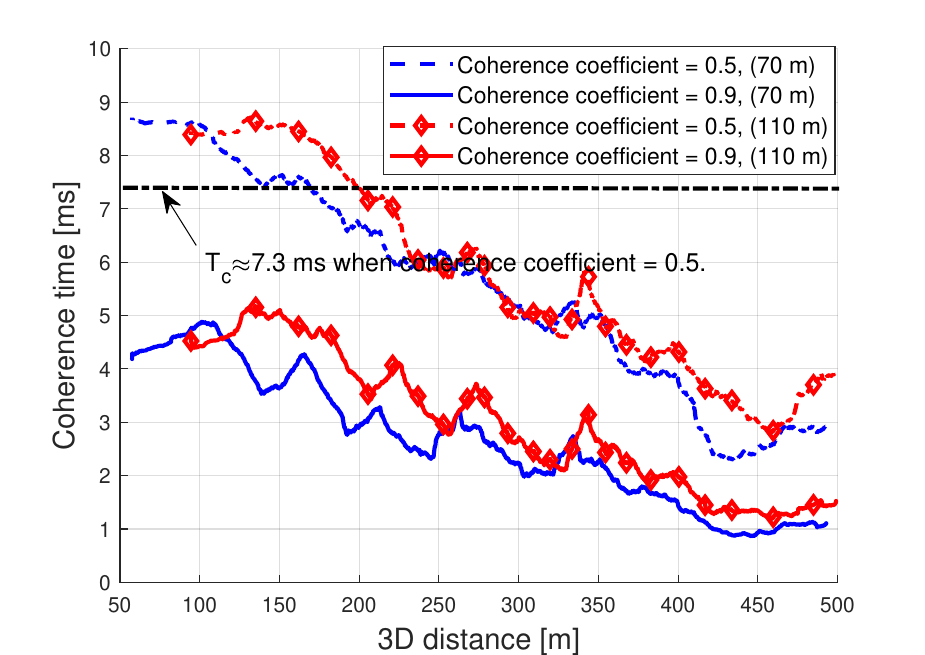}
	\caption{The coherence time depends on the 3D distance and the flight height.}\label{fig:coh_t}
\end{figure}
We calculate the coherence bandwidth and coherence time at  different UAV altitudes using the estimated channels in Section~\ref{sec:chan_est} and evaluate their dependency on link distance. Fig.~\ref{fig:coh_bw} shows the change in the coherence bandwidth. We obtain the coherence bandwidth every 20~ms and rearrange it according to the 3D distance. Then, we apply a moving average filter with a window size of 1000 samples. Coherence bandwidth is calculated by finding the frequency shift where the frequency correlation coefficient becomes 0.9. The frequency correlation coefficient is calculated as
\begin{align}
    R(\Delta f)&=\frac{1}{M}\sum_{i=1}^{M}H(f_i)H(f_i+\Delta f)^{\star},
\end{align}
where $H(f)$ is the frequency domain channel response, $\Delta f$ indicates the frequency shift, and $M$ is the number of available samples. The obtained frequency correlation coefficient $R(\Delta f)$ is normalized by $R(0)$. 

Since we configure 72 subcarriers with 15~kHz subcarrier spacing, the maximum coherence bandwidth is clipped by 1.08~MHz. We observe that the coherence bandwidth decreases as the 3D distance increases. This is because the effect of the multi-path coming from the ground reflection and other objects become stronger as the distance increases. It is also observed that the coherence bandwidth is slightly larger at higher altitude, especially for small and large link distance regions for this particular experiment.

In a similar manner, we analyze the coherence time in Fig.~\ref{fig:coh_t} for two different UAV altitudes. We calculate the coherence time using a segment of 10~ms from the estimated channel. Therefore, the maximum coherence time is clipped to 10~ms. After we obtain coherence time every 20~ms, we rearrange it in ascending order of distance. Then, we use the a moving average window of size  1000 samples. We consider two different  time correlation thresholds for estimating the coherence time, 0.5 and 0.9, respectively. In particular, the time correlation coefficient is calculated as
\begin{align}
    R(\Delta t)&=\frac{1}{M}\sum_{i=1}^{M}H(t_i)H(t_i+\Delta t)^{\star},
\end{align}
where $H(t)$ denotes time domain channel response and $\Delta t$ is time shift. In a single 10~ms channel response, 140 OFDM symbols are included (14 OFDM symbols per 1 subframe), and the time duration of a single OFDM symbol is ${\rm T}_{\rm sym}=71.31$~us. Therefore, we calculate the time correlation coefficient with $\Delta t$ as multiplies of ${\rm T}_{\rm sym}$. We also calculate the coherence time by the generally used approximated formula, and compare it with our results from the measurement.  In particular,  coherence time of the coherence coefficient $= 0.5$ can be theoretically derived by ${\rm T}_{\rm c}\approx\sqrt{\frac{9}{16\pi}}\frac{c}{vf_{\rm c}}=7.3$~ms where $c$, $v$, $f_{\rm c}$ denote speed of light, the velocity of the drone, and carrier frequency, respectively. We use the velocity $v=5$~m/s from Fig.~\ref{fig:UAV_speed}. It is observed that the theoretical coherence time is aligned with the range of coherence time obtained from the measurement results. It is also observed that coherence time decreases  linearly as the 3D distance increases for both 70~m and 110~m UAV heights. In addition,  coherence time is slightly larger at 110~m when compared with the 70~m UAV height, for both coherence coefficients of 0.5 and 0.9.

\section{Shadowing Distribution and Spatial Correlation Analysis}\label{sec:spatial_cor}
\begin{figure}[t]
	\centering
	\includegraphics[width=0.48\textwidth]{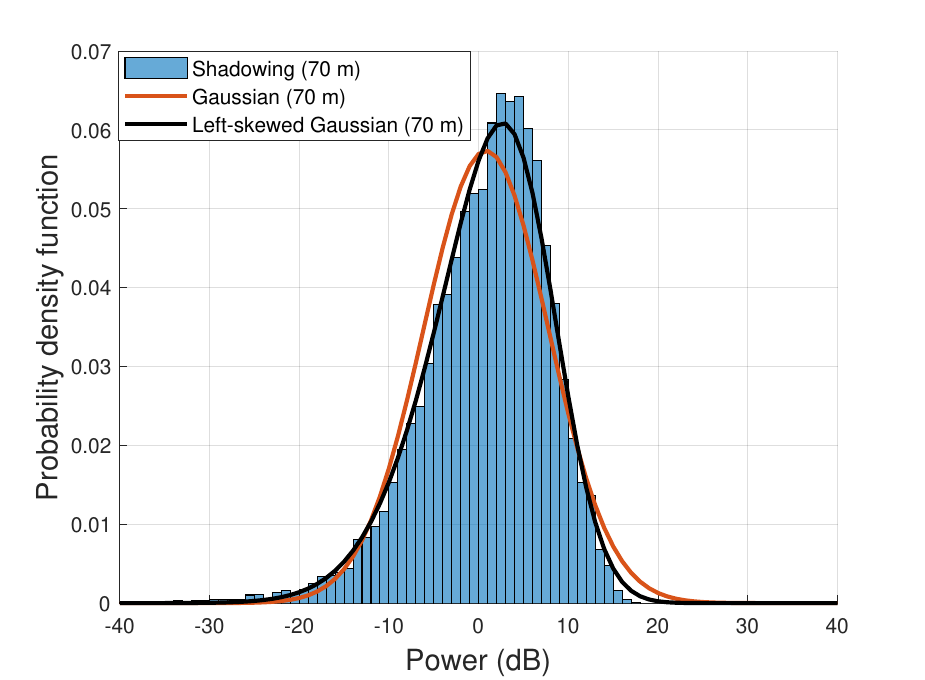}
	\caption{Probability density function of the shadowing component at 70~m UAV height where it is fitted by Gaussian and left-skewed Gaussian where $\alpha=-2$ in \eqref{eq:skew_Gau}.}\label{fig:shad_dist}
\end{figure}
\begin{figure}[t]
	\centering
	\subfloat[Correlation of horizontal distance]{\includegraphics[width=0.48\textwidth]{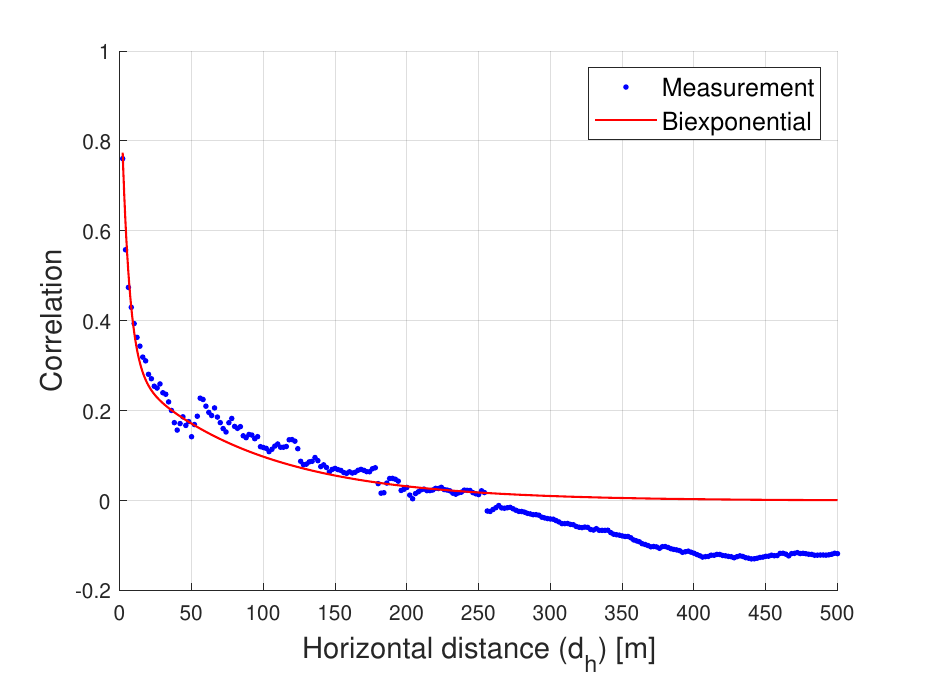}}
 
    \subfloat[Correlation of vertical distance]{\includegraphics[width=0.48\textwidth]{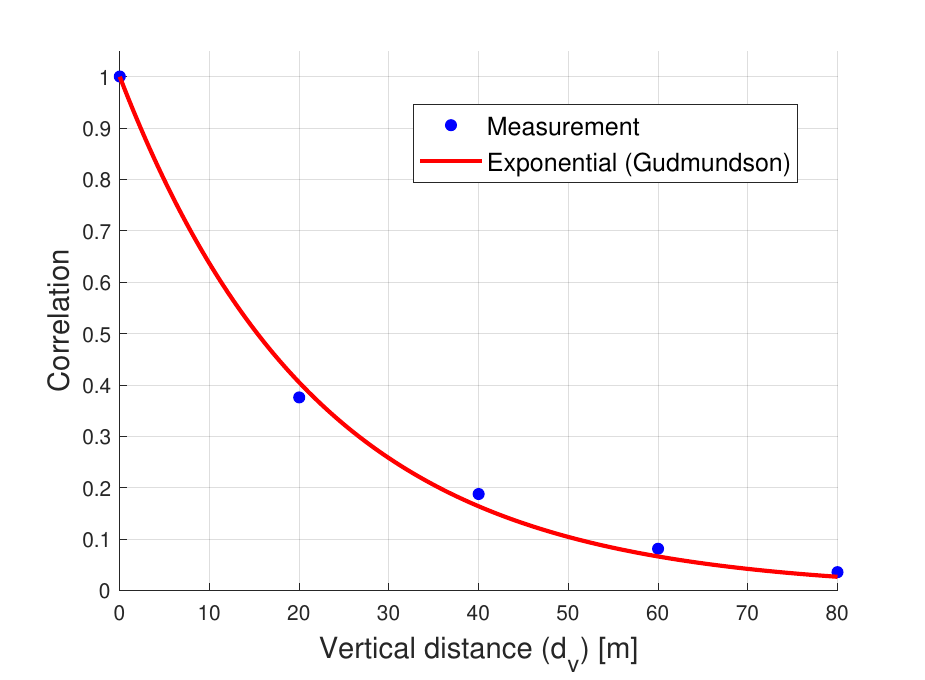}}
	\caption{Correlation of horizontal and vertical distance where it is fitted by bi-exponential and exponential functions where fitting parameters $a=0.3$, $b_1=0.0112$, $b_2=0.188$ in \eqref{eq:corr_dh}.}\label{fig:corr}
\end{figure}
\begin{figure}[t]
	\centering
	\includegraphics[width=0.48\textwidth]{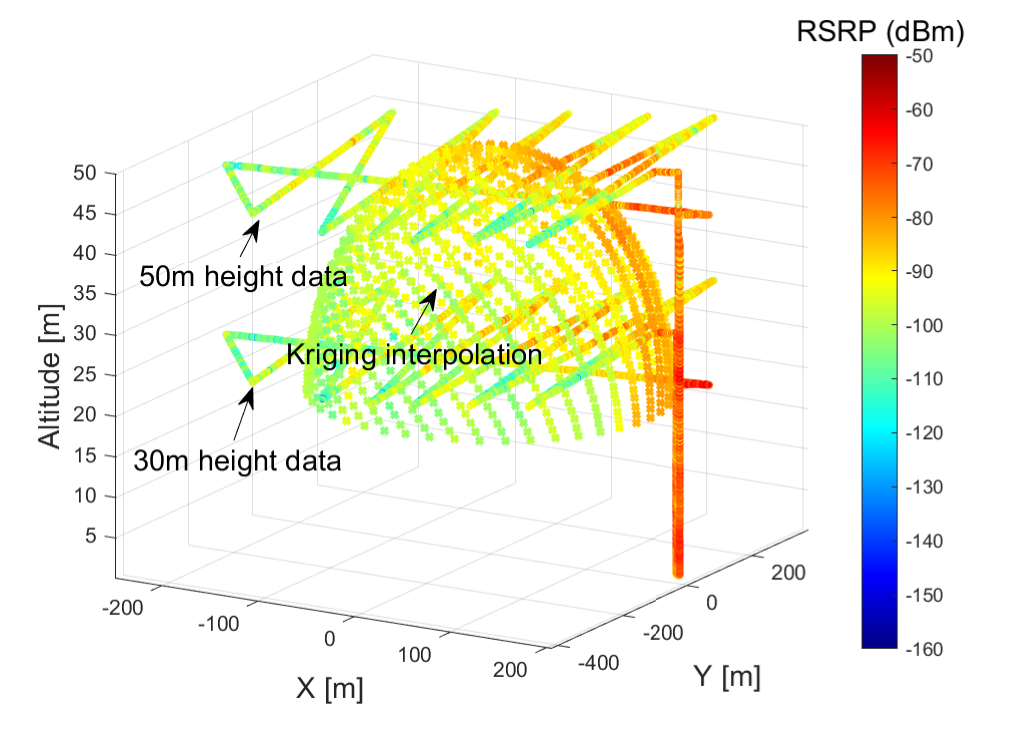}
	\caption{RSRP interpolation by Kriging at the surface of a 3D volume. Measurement RSRP of 30~m and 50~m altitudes are utilized.}\label{fig:Kriging_example}
\end{figure}
In this section, we analyze shadowing and spatial correlation in terms of horizontal distance and vertical distance. We obtain the shadowing component from measured RSRP by subtracting Tx power and two-ray path loss, which is obtained from Section~\ref{sec:RSRP}. Fig.~\ref{fig:shad_dist} shows the probability density function (PDF) of the shadowing component which is fitted by Gaussian and left-skewed Gaussian distributions. It is observed that the shadowing component is better modeled using a left-skewed Gaussian distribution rather than a Gaussian distribution. The PDF of the skewed Gaussian distribution is given by
\begin{align}\label{eq:skew_Gau}
    f(x)=2\phi\left(\frac{x-\xi}{\omega}\right)\Phi\left(\alpha\left(\frac{x-\xi}{\omega}\right)\right),
\end{align}
where $\phi(\cdot)$ and $\Phi(\cdot)$ indicate the PDF and the cumulative distribution function (CDF) of Gaussian distribution, respectively. The parameter $\alpha=-2$ in \eqref{eq:skew_Gau} decides the skewness of the distribution.

We also analyze the spatial correlation as a function of the horizontal and vertical distance between the locations that the RSRP is calculated. Since height is fixed throughout a certain experiment, and the horizontal trajectory is identical for experiments at different UAV altitudes, we can separately generate vertical and horizontal correlation. The correlation between two different location samples $w_{i}$, $w_{j}$ is calculated by
\begin{align}
    R_{i,j}=\frac{(w_{i}-\nu_i)(w_{j}-\nu_j)}{\sigma_{w,i}\sigma_{w,j}},
\end{align}
where $\nu$, $\sigma_{w}$ denote the mean and the standard deviation of the samples. Fig.~\ref{fig:corr} shows the correlation from the measurements, along with the bi-exponential function fit for horizontal distance correlation and  the exponential function fit for the vertical distance correlation. The exponential function correlation model is also known as the Gudmundson model~\cite{gudmundson1991correlation}.
The bi-exponential function is expressed as
\begin{align}\label{eq:corr_dh}
    R(d_{\rm h})=ae^{-b_1d_{\rm h}}+(1-a)e^{-b_2d_{\rm h}},
\end{align}
where $a=0.3$, $b_1=0.0112$, $b_2=0.188$ are the fitting parameters.

The measured RSRP at multiple altitudes can be utilized to generate a radio map of a 3D volume area or surface. The radio map of an area can monitor and capture signal leakage from a RDZ. We interpolate measured RSRP with GPS information using Kriging interpolation technique. Kriging can predict the signal strength of unknown locations by the combination of the signal strength of nearby known locations using a semi-variogram which can be obtained from spatial correlations in Fig.~\ref{fig:corr}. In Fig.~\ref{fig:Kriging_example}, we show a representative result of RSRP interpolation of an area where the RSRP at the surface of a dome shape volume is interpolated by 30~m and 50~m measurement datasets.   

\section{Conclusion}\label{sec:conclusion}
In this paper, we analyze the channel propagation of air-to-ground LTE links using an experimental dataset of raw LTE I/Q samples at 3.51 GHz. We conduct a measurement campaign by using SDR and GPS receivers mounted on UAVs in a rural environment. The UAV sweeps across the experimental area following the predefined fixed-height trajectories. The I/Q data is collected at altitudes ranging from 30~m to 110~m, at identical horizontal trajectories. We process the measured I/Q samples by using MATLAB LTE Toolbox and analyze the synchronization, cell search, channel estimation, and RSRP measurement procedures. We post-process the received signal power by fitting it with a two-ray path loss model considering the Tx and the Rx antenna patterns. We compare and analyze the results at different UAV heights to observe the height-dependent channel characteristics. Our future work includes repeating a similar study using the I/Q data from 5G wireless transmissions. 

\section*{Acknowledgement}
This research is supported in part by the NSF award CNS-1939334 and its associated supplement for studying NRDZs. The measurement I/Q sample datasets used in this work and post-processing codes are available in \cite{AERPAW_NRDZ_site}.

\bibliographystyle{IEEEtran}
\bibliography{IEEEabrv,references}
  
\end{document}